\documentclass[aps,preprint]{revtex4-1}
\usepackage[french,english]{babel}
\usepackage{amsmath,amssymb}
\usepackage{graphicx}
\usepackage{verbatim}
\usepackage{color}
\usepackage{hyperref}
\usepackage{eqnarray,tabularx}
\usepackage{ccaption,float,esvect,amsthm}
\usepackage{wrapfig}
\usepackage{amsbsy,amssymb,bm}
\usepackage{pstricks,pst-circ,pstricks-add,pst-node,pst-coil,pst-infixplot,pst-eucl,pst-text,pst-grad,pst-fun,pst-diffraction,pst-3d,pst-tools}
\usepackage{amsmath}
\usepackage{subfigure,xcolor}

\newtheorem{theorem}{Theorem}
\newtheorem*{Proof}{Proof}
\newtheorem*{consequence}{Consequence}

\begin{document}
\title{ Yet another approach to the inverse square law and to the circular character of the hodograph of Kepler orbits}
\author{Adel H. Alameh}
\affiliation{Lebanese University, Department of Physics, Hadath, Beirut, Lebanon}
\date{\today}
\email{adel.alameh@eastwoodcollege.com}

\begin{abstract}

\noindent The law of centripetal force governing the motion of celestial bodies in eccentric conic sections, has been established and thoroughly investigated by Sir Isaac Newton in his Principia Mathematica. Yet its profound implications on the understanding of such motions is still evolving.\\

\noindent In a paper to the royal academy of science, Sir Willian Hamilton demonstrated that this law underlies the circular character of  hodographs for Kepler orbits. A fact which was the object of ulterior research and exploration by Richard Feynman and many other authors~\cite{history}.\\

\noindent  In effect, a  minute examination of  the geometry of elliptic trajectories,  reveals interesting geometric properties and relations, altogether, combined  with the law of  conservation of angular momentum lead eventually, and without any recourse to dealing with differential equations, to the appearance of the equation of the trajectory and to the  derivation of the equation of its corresponding hodograph.\\

\noindent  On this respect, and for the sake of founding the approach on  solid basis, I devised  two mathematical theorems; one concerning the existence of geometric means, and the other is related to establishing the parametric equation of an off-center circle, altogether compounded with other simple arguments ultimately give rise to the inverse square law of force that governs the motion of bodies in elliptic trajectories, as well as  to the  equation of their inherent circular hodographs.

\end{abstract}
\maketitle
\section*{Preliminary geometry of elliptic trajectories}
\begin{figure}[H]
\centering
\begin{minipage}{.8\linewidth}
\includegraphics[]{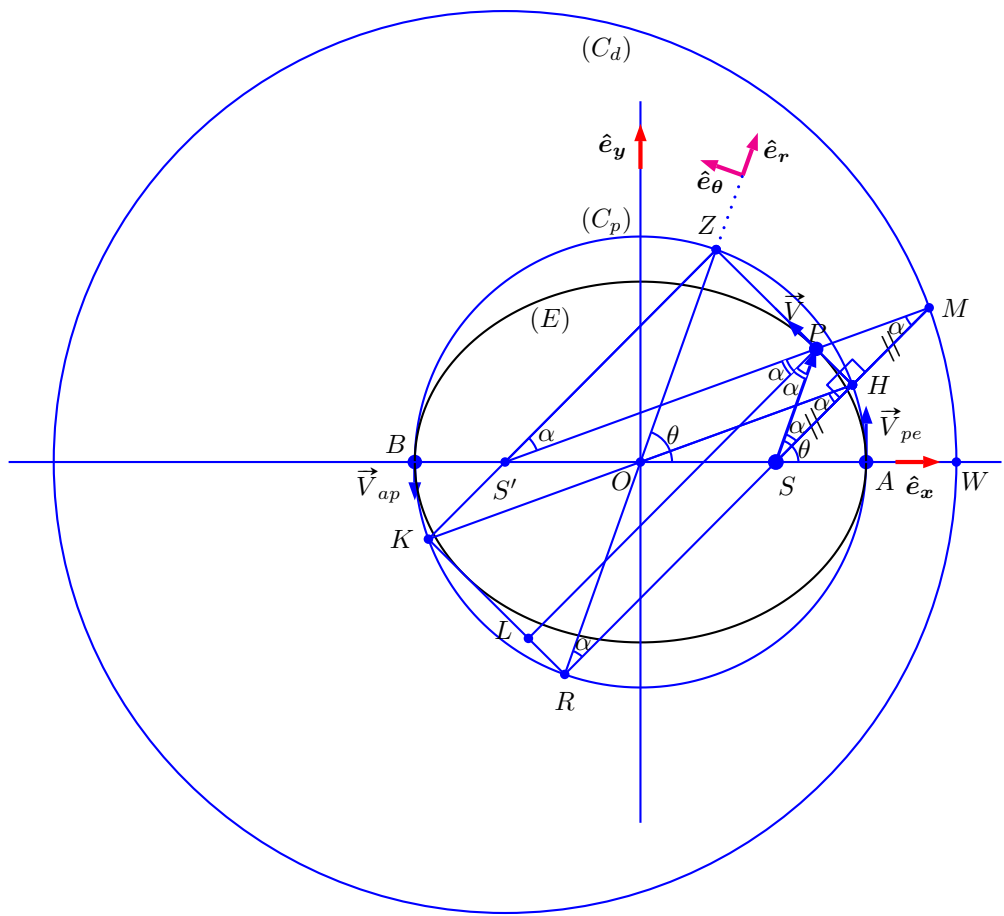}
\end{minipage}

\caption{Geometry of elliptic orbits}
\label{geometry}

\end{figure}

\noindent Let  $S$ and $S'$,  separated by $SS'=2c$, be the foci of an ellipse $(E)$, described by a celestial body  $P$ (a planet), that is  impelled by a force tending toward a center of force $S$ (a star). Figure~\ref{geometry}. Draw the principal circle $(C_p)$ of center $O$ and radius $OA=a$, where $a$ is the length of the semi major axis of $(E)$. Draw the director circle $(C_d)$ of center $S'$ and radius $S'M=2a$. Produce $S'P$ to meet the director circle $(C_d)$ in $M$, let fall the perpendicular $PH$ to $SM$, then  produce $HP$ to  meet the principal circle $(C_p)$ in $Z$. Produce $MS$ to meet $(C_p)$ in $R$.

 It is now required to prove that the radius vector $\bm{r}=\bm{SP}$ is parallel to $\bm{RZ}$\,. For that purpose, we start off with  an alternative definition of the ellipse~\cite{ellipse},  which states that an ellipse  is the locus of the centers of circles passing through the focus $F$ and internally tangent to the director circle whose center lies at the other focus $F'$ and of radius $2a$. The case being so, it is easy to infer that $\Delta\, PSM$ is isosceles, and that $H$ is the midpoint of $SM$.
Let $PL$ be the bisector of the angle $\widehat{SPS'}$. We seek now to prove that $PL$ is parallel to $SM$. Evidently $\widehat{MPH}=\widehat{ZPS'}$ as they are vertically opposite angles, and
$\widehat{MPH}=\widehat{SPH}$ corresponding parts of congruent triangles, hence $\widehat{ZPS'}=\widehat{HPS}$, and it is readily inferred that $\widehat{HPL}=90^\circ$. Hence, $PL$ is parallel to $SM$ and correlatively $PH$ is tangent to the ellipse since it is perpendicular to $PL$ which is the internal bisector of angle $\widehat{SPS'}$\,.\\

\noindent Now in  $\Delta\, S'SM$, the points $O$ and $H$ are the midpoints of $S'S$ and $SM$ respectively, then $OH=\displaystyle\frac{S'M}{2}=a$, therefore $H$ belongs to the principal circle $(C_p)$, and having the angle $\widehat{RHZ}=90^\circ$, leads to saying that   $ZR$ is a diameter of the principal circle.\\
Few more steps are still needed to achieve the required proof, thus we proceed  by observing that $\widehat{PSH}=\widehat{PMH}$ since $\Delta\, PSM$ is isosceles,
   furthermore, $\Delta\, ORH$ is also isosceles, for $OR=OH$ radii of the same circle. Hence $\widehat{ORH}=\widehat{OHR}$\,; but $\widehat{OHR}=\widehat{PMH}$, since $OH$ is parallel to $S'M$, consequently  $\widehat{OHR}=\widehat{PMH}$, thus
 $\widehat{PSH}=\widehat{ORH}$. And correlatively $SP$ is parallel to $RZ$\,,
 therefore $\widehat{PSW}=\widehat{ZOS}=\theta$.

\section*{The geometric mean theorem}
\begin{theorem} Let $f\colon x\rightarrow f(x)$ be a continuous function that does not vanish anywhere on the interval $[a,b]$ and differentiable within it,
such that $f^{\prime}(x)\neq 0$ within the interval, then there exist
a point of abscissa $x=c$\,,\,\,\,where\,\,\,$a<c<b$\,\, such that

\begin{equation}f^{2}(c)=f(a)\cdot f(b)\label{h}\end{equation}
\end{theorem}
\begin{Proof}
Let us construct the auxiliary function $\beta(x)$ defined in the interval $[a,b]$ such that
\begin{equation}\beta(x)=\displaystyle\frac{f(x)}{f(a)}+\displaystyle\frac{f(b)}{f(x)}\end{equation}
Now
\begin{equation}\beta(a)=1+\displaystyle\frac{f(b)}{f(a)}\end{equation}
And
\begin{equation}\beta(b)=\displaystyle\frac{f(b)}{f(a)}+1\end{equation}
Therefore
\begin{equation}\beta(a)=\beta(b)\end{equation}
And so, by {{\it Rolle's theorem}}~\cite{Rolle} there exist a point of abscissa $x=c$\,,  such that $\beta^{\prime}(c)=0$
\begin{equation}\beta^{\prime}(c)=\displaystyle\frac{f^{\prime}(c)}{f(a)}-\displaystyle\frac{f(b)\cdot f^{\prime}(c)}{f^{2}(c)}=0 \end{equation}
And by rearranging and canceling $f^{\prime}(c)$ we get
\begin{equation}f^{2}(c)=f(a)\cdot f(b)\end{equation}
And the theorem is proved.
\end{Proof}
\section*{Probing into Kepler elliptic trajectories}
\noindent The radius vector $\bm{r}$ of a planet moving on an elliptic orbit, is a continuous function of the angle $\theta$ that it makes with the major axis from the perihelion side. The angle $\theta$  in  turn changes also with time. The modulus of $\bm{r}$ takes a minimum value $r_p=a-c$ at $\theta=0$, when the planet passes through the perihelion and a maximum value $r_a=a+c$ at $\theta=\pi$, in its passage through the aphelion, then according to  theorem (1), there exist a value $\theta_b$  of $\theta$  such that $r^2(\theta_b)=r(0) r(\pi)$, that is  $r^2(\theta_b)=a^2- c^2$. But in the case of an ellipse, the length of the semi minor axis is given by the relation  $b^2=a^2-c^2$. Therefore the modulus of the radius vector should take a value equals to the length of the semi minor axis of the ellipse at a well specified angle. I called it $\theta_b$.

\noindent Returning to the geometric features of the figure~\ref{geometry}, by a well known relation we have$\colon$\\
\begin{equation}SH\times SR=SA\times SB \end{equation}
that means
\begin{equation}SH\times SR=(a-c)(a+c)=b^2~\label{principalcircle1}\end{equation}
The position vector of the planet expressed in polar coordinates has the form $\bm{r}=r\,{\bm{\hat{e}_r}}$ and its velocity vector is $\bm{V}=\displaystyle\frac{d\bm{r}}{dt}=\dot{r}\,{\bm{\hat{e}_r}}+r\dot\theta\,{\bm{\hat{e}_\theta}}$\,.
Now, since the planet is urged by a centripetal force towards the star, so the applied external torque on the planet is zero, hence its angular momentum $\bm{L}=\bm{r}\times m\bm{V}$ is conserved~\cite{Goldestein}. The magnitude of the angular momentum is
$L=mV_\theta r$ where $V_\theta=r\dot\theta$ is the transverse component of the velocity vector. The magnitude of $L$ may also be expressed as $L=m\times V\times SH$ owing to  the fact that $SH=r\,\sin\theta$.\\ So, we can say that$\colon$
\begin{equation} L=mV_\theta r=m V\times SH \end{equation}
 By canceling $m$\,, we get the expression of $h$ which is that of the angular momentum per unit mass as being$\colon$
\begin{equation}h=rV_\theta=SH\times V=r^2\dot\theta ~\label{momentum}\end{equation}
 In the course of motion, the radius vector should, at a certain moment, take the value $r=b$ at the angular position $\theta=\theta_b$, accordingly $h$ being constant can be expressed as $h=b^2\dot\theta_b$\,, where $\dot\theta_b$ is the angular velocity at $\theta=\theta_b$.
So, from equation~(\ref{momentum}) we can say that
\begin{equation} SH\times V=b^2\,\dot\theta_b ~\label{beem}\end{equation}
and
\begin{equation}r\,V_\theta=b^2\,\dot\theta_b ~\label{beemm}\end{equation}
Now multiplying equation~(\ref{principalcircle1}) by $\dot\theta_b$ we get
\begin{equation}SH\times SR\times\dot\theta_b=b^2\dot\theta_b~\label{principalcircle2}\end{equation}
 And by comparing equations~(\ref{beem}) and~(\ref{principalcircle2}) and  we get $V=SR\times \dot\theta_b$\,.\\
But, $SR=ZS'$, since $\Delta\, OS'Z$ is equal to $\Delta \,OSR$, for $OS=OS'=c$, $OZ=OR=a$ and $\widehat{S'OZ}=\widehat{SOR}$ vertically opposite angles, hence,
\begin{equation} V=ZS'\times \dot\theta_b \end{equation}

\noindent Therefore \begin{equation} \bm{V}=ZS'\times\dot\theta_b \,\bm{\hat{e}_t}~\label{ange}\end{equation}
Where $\bm{\hat{e}_t}$ is a unit vector tangent to the trajectory in the same direction as $\bm{V}$.\\
\emph{In the language of mathematics, the velocity vector $\bm{V}$ is the image of $\bm{ZS'}$ by a direct similitude of ratio $\dot\theta_b>0$ and of an angle $(\bm{ZS'},\bm{V})=-\displaystyle\frac{\pi}{2} +2k\pi$.}

\noindent It remains only to put equation~(\ref{ange}) in a more explicit form, by finding the expressions of $S'Z$ and $\bm{\hat{e}_t}$ in terms of the dynamic parameters of the motion. For that purpose and returning to figure~\ref{geometry}, we notice that $\bm{S'Z}=\bm{SO} +\bm{OZ}$ i.e. $\bm{S'Z}=c\,\bm{\hat{e}_x}+ a\,\bm{\hat{e}_r}$ since $\bm{OZ}$ is parallel to $\bm{r}$ as proved before. But $\bm{\hat{e}_r}=\cos\theta \,\bm{\hat{e}_x} + \sin\theta\,\bm{\hat{e}_y}$, therefore, $\bm{S'Z}=(c+a\cos\theta)\,\bm{\hat{e}_x}+a\sin\theta\,\bm{\hat{e}_y}$\,.
Then expressing it in  polar coordinates by the well known transformation relation$\colon$
\begin{equation}\left(\begin{array}{c}\bm{\hat{e}_x}\\\bm{\hat{e}_y}\\ \end{array}\right)=\left(\begin{array}{lr}\cos\theta&-\sin\theta\\\sin\theta&\cos\theta\\\end{array}\right)
\left(\begin{array}{c}\bm{\hat{e}_r} \\ \bm{\hat{e}_\theta} \end{array}\right)~\label{matrix}\end{equation}

\noindent We get $\bm{S'Z}$ to be$\colon$
\begin{equation}\bm{S'Z}=(a+c\cos\theta)\,\bm{\hat{e}_r}-c\sin\theta\, \bm{\hat{e}_\theta}\end{equation}
Thus  the modulus of $\bm{S'Z}$ will have the expression
\begin{equation} [S'Z]=\sqrt{a^2+c^2+2ac\cos\theta}~\label{moduluss}\end{equation}
and a unit vector $\bm{\hat{e}_n}$ along the normal to the trajectory should be
$\bm{\hat{e}_n}=-\displaystyle\frac{\bm{S'Z}}{[S'Z]}$, So
\begin{equation}\bm{\hat{e}_n}=-\displaystyle\frac{a+c\,\cos\theta}{\sqrt{a^2+c^2+2ac\,\cos\theta}}\,\bm{\hat{e}_r}+\displaystyle\frac{c\,\sin\theta}
{\sqrt{a^2+c^2+2ac\,\cos\theta}}\,\bm{\hat{e}_\theta}\end{equation}
The unit vector $\bm{\hat{e}_t}$ that is in the direction of the velocity vector $\bm{V}$ is perpendicular to $\bm{\hat{e}_n}$ and its expression will be
\begin{equation} \bm{\hat{e}_t}=+\displaystyle\frac{c\,\sin\theta}{\sqrt{a^2+c^2+2ac\,\cos\theta}}\,\bm{\hat{e}_r}+\displaystyle\frac{a+c\,\cos\theta}{\sqrt{a^2+c^2+2ac\,\cos\theta}}\,
\bm{\hat{e}_\theta}~\label{tangent1}\end{equation}

\noindent  Finally by substituting~(\ref{tangent1}) and (\ref{moduluss}) in~(\ref{ange}) we obtain the expression of the velocity vector
\begin{equation}\bm{V}= c\,\dot\theta_b\,\sin\theta\,\bm{\hat{e}_r}\,+\,(a+c\,\cos\theta)\,\dot\theta_b\,\bm{\hat{e}_\theta} ~\label{velocity}\end{equation}

\noindent Accordingly we pursue our search to retrieve the equation of the elliptic trajectory from what preceded. It is to be noticed in this context that,
 the components of the velocity vector in the polar system are given by equation~(\ref{velocity}) as$\colon$
 \begin{equation}V_r=c\,\dot\theta_b\sin\theta~\label{vr}\end{equation}
 \begin{equation}V_\theta=(a+c\cos\theta)\,\dot\theta_b~\label{vtheta}\end{equation}
Substituting $V_\theta$ as given by equation~(\ref{vtheta}) in equation~(\ref{beemm}) and canceling $\dot\theta_b$\,, will give rise to the  equation of the ellipse in a harmonious fashion.

\begin{equation} r=\displaystyle\frac{b^2}{a+c\cos\theta}=\displaystyle\frac{a(1-\epsilon^2)}{1+\epsilon\cos\theta}\end{equation}
Where $\epsilon=\displaystyle\frac{c}{a}$ is the eccentricity of the ellipse.\\
Further, the expression of $\bm{V}$ can also be transformed into the cartesian system by the use of equation~(\ref{matrix}) in the inverted form, thus$\colon$
\begin{equation}\bm{V}=-a\,\dot\theta_b\,\sin\theta \, \bm{\hat{e}_x} +(c+a\,\cos\theta)\,\dot\theta_b \,   \bm{\hat{e}_y} ~\label{velocitycartesian}\end{equation}
and its modulus will be
\begin{equation}V=\dot\theta_b\,\sqrt{a^2+c^2+2ac\,\cos\theta}~\label{modulus}\end{equation}
and by squaring equation~(\ref{modulus}) we get
\begin{equation} V^2\,=\,(a\,\dot\theta_b)^2\,+\,2(a\,c)\,\dot\theta_b^2\,\cos\theta \,+\,(c\,\dot\theta_b)^2\end{equation}
then by calling $V_c=a\,\dot\theta_b$ and $V_0=c\,\dot\theta_b$ we obtain
\begin{equation} V^2\,=\,V_c^2\,+2\,V_c\,V_0\,\cos\theta\,+\,V_0^2 ~\label{modulus1}\end{equation}

 It must be pointed out  that equation~(\ref{modulus1}) has a major importance in  providing the value of $V$ in terms of the angle $\theta$ that the planet makes with the perihelion at any moment.
\noindent One can also add a peculiar privilege to this approach among others yet to come, namely that of equation~(\ref{ange}) that introduced the second focus of the ellipse into the scene.\\
  Moreover, it is to be noticed that if we plug $\theta=0$ at the perihelion and $\theta=\pi$ at the aphelion into equations~(\ref{vr}) and (\ref{vtheta}), one would obtain that in both cases the radial velocity is zero and that the transverse components of the velocity vector $\bm{V}$ will take respectively the values $V_{pe}=(a+c)\,\dot\theta_b$ and $V_{ap}=(a-c)\,\dot\theta_b$. But $V_{pe}=(a-c)\,\dot\theta_{pe}$ and $V_{ap}(a+c)\,\dot\theta_{ap}$\,. Then by matching the two expressions of $V_{pe}$ and $V_{ap}$, one obtains$\colon$
   \begin{equation}\dot\theta_{pe}=\displaystyle\frac{a+c}{a-c}\,\dot\theta_b\,\,\,\,\,\,\,\,\mathrm{and} \,\,\,\,\,\,\,\, \dot\theta_{ap}=\displaystyle\frac{a-c}{a+c}\,\dot\theta_b ~\label{ceta}\end{equation}
 Finally by multiplying the last two expressions we get
\begin{equation}\dot\theta_b^2=\dot\theta_{pe}\,\dot\theta_{ap}~\label{mean}\end{equation}
and it turns out that our $\dot\theta_b$ is nothing but the geometric mean of $\dot\theta_{pe}$ and $\dot\theta_{ap}$\,.\\
We now turn our attention to deriving the law of force, so we proceed by rearranging equation~(\ref{velocity}) to the form
\begin{equation}\bm{V}=c\,\dot\theta_b\,(\sin\theta\,\bm{\hat{e}_r}+\cos\theta\,\bm{\hat{e}_\theta})+a\,\dot\theta_b\, \bm{\hat{e}_\theta} \end{equation}
and knowing from equation~(\ref{matrix}) that $\bm{\hat{e}_y}=\sin\theta\,\bm{\hat{e}_r}+\cos\theta\,\bm{\hat{e}_\theta}$ we obtain
\begin{equation}\bm{V}=c\,\dot\theta_b\,\bm{\hat{e}_y}+a\,\dot\theta_b \,\bm{\hat{e}_\theta}~\label{velocityy}\end{equation}
 then we derive equation~(\ref{velocityy}) with respect to time, and knowing that $\displaystyle\frac{d\bm{\hat{e}_y}}{dt}=\bm{0}$\,,\\ and $\displaystyle\frac{d\bm{\hat{e}_\theta}}{dt}=-\dot\theta\,\bm{\hat{e}_r}$ we thus obtain the expression of the acceleration vector
\begin{equation} \bm{\mathcal{A}}=-\,a \,\dot\theta_b\, \dot\theta \,\bm{\hat{e}_r}~\label{acc}\end{equation} and given that   $r^2\,\dot\theta=b^2\,\dot\theta_b$ means that $\dot\theta$ can be expressed as  $\dot\theta=\displaystyle\frac{b^2\,\dot\theta_b}{r^2}$ which when substituted in equation~(\ref{acc}), gives the expression of the acceleration vector as

\begin{equation}\bm{\mathcal{A}}=-\displaystyle\frac{a\,b^2\,\dot\theta_b^2}{r^2}\,\bm{\hat{e}_r}\end{equation}
And on the basis of Newton's second law $\bm{\mathcal{F}}\, =\,m\,\bm{\mathcal{A}}$\,, one obtains
\begin{equation}\bm{\mathcal{F}}=-\displaystyle\frac{m\,a\,b^2\,\dot\theta_b^2}{r^2}\,\bm{\hat{e}_r}\end{equation}
and knowing that $b=a\sqrt{1-\epsilon^2}$ we get
\begin{equation}\bm{\mathcal{F}}=-\displaystyle\frac{m\,a^3\,(1-\epsilon^2)\,\dot\theta_b^2}{r^2}\,\bm{\hat{e}_r}~\label{Newton}\end{equation}
In Newton's law of universal gravitation, the force is given as$\colon$
\begin{equation} \bm{\mathcal{F}}=-\displaystyle\frac{GmM}{r^2}\,\bm{\hat{e}_r}~\label{Newton1}\end{equation}
and by comparing equations~(\ref{Newton}) and~(\ref{Newton1}) one obtains the value of $\dot\theta_b$ to be \begin{equation}\dot\theta_b=\sqrt{\displaystyle\frac{GM}{a^3(1-\epsilon^2)}}~\label{thetadot}\end{equation}
Moreover, it appears that a closer inspection of equation~(\ref{thetadot}), and given that the star is permanently consuming its mass in favor of energy of electromagnetic radiations and other particles, then one  might infer that it  had at an earlier stage a mass
\begin{equation}M'=\displaystyle\frac{M}{1-\epsilon^2}~\label{mass}\end{equation}
and then the  angular velocity of the planet would have been $\dot\theta_b=\sqrt{\displaystyle\frac{GM'}{a^3}}$ which is that of a uniform circular motion.  Accordingly, one could infer that the planet should have been revolving at that stage in a uniform circular motion, and hence its orbit is becoming more and more elliptic with time.

\noindent Furthermore, equation~(\ref{mass}) gives rise to the law that governs the variation of the eccentricity of the elliptic orbit with time as follows
\begin{equation} \epsilon(t)=\sqrt{1-\displaystyle\frac{M(t)}{M'}}~\label{eccentricity}\end{equation}

\noindent A knowledge of the actual power of the star, along with the use of Einstein's mass energy relation $E=mc^2$ would not be enough to find exactly the relation between the current mass $M$ of the star and its earlier mass $M'$ when the planet was orbiting it in uniform circular motion, because a part of the mass of the star flees it randomly through stellar winds. Despite these difficulties, an interesting feature may be extracted from equation~(\ref{eccentricity})  namely that of the effect of the mass on the geometry, as it indicates that the geometry of the orbit is changing from circular to elliptic as the mass of the star decreases with time. On this respect, one should also notice that these spontaneous modifications in the geometry of the orbit occur in a sense as to change the orbit from the most ordered shape (circle), to a less ordered shape (ellipse). \\
 Another implication of Newton's law of gravitational interaction expressed in the form of equation~(\ref{Newton}) may be noticed when it comes to a moon of mass $m$ orbiting a planet of presumably constant mass $M$\,, on an elliptic orbit, then one could predict that another moon of mass \mbox{$m'=\,m\,(1-\epsilon^2)$}\,, let go with the same initial conditions as $m$\,, would orbit the planet in a uniform circular motion, because its angular velocity would then have been $$\dot\theta_b=\sqrt{\displaystyle\frac{GM}{a^3}}$$

 \begin{consequence}
  A planet orbiting a star in an elliptic orbit should possess at two specific instants  every one complete revolution around the star  an angular velocity that is equal to its angular velocity had it been rotating in a uniform circular motion at an earlier stage of the life of the star.
  \end{consequence}

\noindent Before we proceed to extract the information from equation~(\ref{modulus1}), a little digression into defining the hodograph is needed. Thus,
the hodograph is the curve generated  by the tip of a vector equipollent to the velocity vector and whose tail lies at the origin of the velocity space.
The velocity vector of a moving body is permanently tangent to the trajectory described by that body at any instant. Except for uniform circular motion in which the modulus of the velocity vector remains constant,  all other sorts of  curvilinear motion  are characterized by a changing velocity vector  regarding modulus and direction. Nonetheless we still have need to construct the equation of an off-center circle in polar coordinates and for that sake we introduce theorem (2).

\begin{figure}[H]
\centering
\begin{minipage}{.4\linewidth}
\includegraphics[]{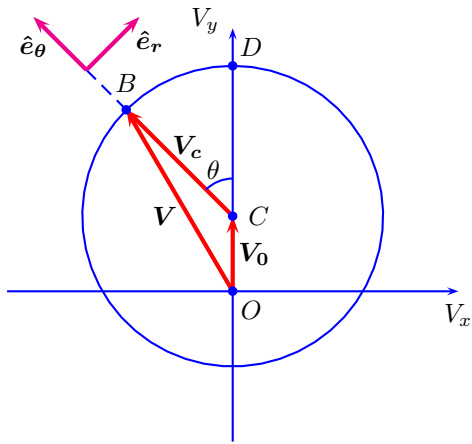}
\end{minipage}
\caption{Hodograph of an elliptic trajectory}
\label{Hodo}
\label{Fig}

\end{figure}

\begin{theorem}
The modulus of the radius vector of a point $M$ moving on a circle of radius $\rho$ centered at $(r_0,\phi_0)$ and of parameter $\theta$  is defined by$\colon$
\begin{equation}r^2=\rho^2 +2\rho \,r_0\cos\theta+r_0^2\end{equation}

\begin{Proof}

Let us consider a circle $\Omega$  (figure~\ref{Relations}) of radius $\rho$ and center $C(r_0,\phi_0)$\,. A point $M$ on the circle is located by its radius $r$ and by the azimuthal angle $\phi$\,.

\begin{figure}[H]
\centering
\begin{minipage}{.8\linewidth}
\includegraphics[]{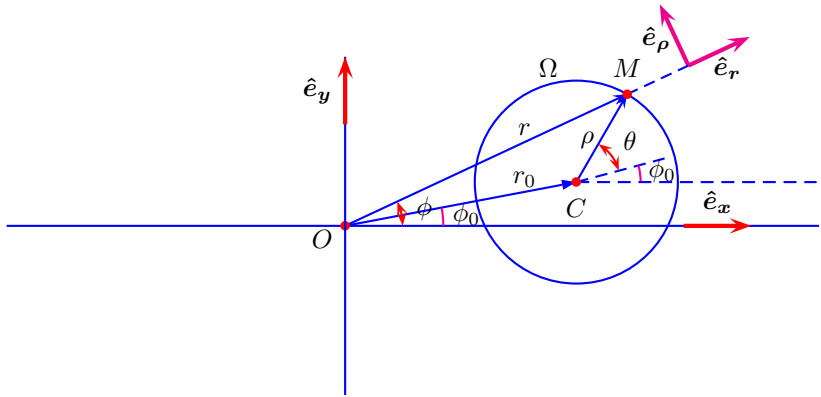}
\end{minipage}
\caption{Relations satisfied by off-center circles in polar coordinates }
\label{Relations}
\end{figure}

Now
 \begin{equation}\bm{r}=\bm{r_0}\,+\,\bm{\rho}\label{circle}\end{equation}
 Projecting equation~(\ref{circle}) successively on $\bm{\hat{e}_x}$ and $\bm{\hat{e}_y}$ we get
 \begin{equation} r\cos\phi=r_0\cos\phi_0+\rho\cos(\theta+\phi_0)\label{circle1}\end{equation}
 and
 \begin{equation}r\sin\phi=r_0\sin\phi_0+\rho\sin(\theta+\phi_0)\label{circle2}\end{equation}
 In reality equations~(\ref{circle1}) and (\ref{circle2}) represent the parametric equations of the circle $\Omega$
 \begin{equation}\left\{\begin{array}{lr}x=\rho_0\cos\phi_0 +\rho\cos(\theta+\phi_0)& ~\\
y=\rho_0\sin\phi_0 + \rho\sin(\theta +\phi_0)& ~\\
\end{array}\right. ~\label{equcircle}\end{equation}

\noindent Then, by squaring and adding equations~(\ref{circle1}) and~(\ref{circle2}) we get
 \begin{equation}r^2=r_0^2+\rho^2+2r_0\rho[\cos(\theta+\phi_0)\cos\phi_0 +\sin(\theta+\phi_0)\sin\phi_0]\end{equation}
 But
 \begin{equation} \cos(m-n)=\cos m \cos n + \sin m \sin n \end{equation}
 Therefore
 \begin{equation}r^2=\rho^2+2\rho\,r_0\cos\theta + r_0^2\label{circle3}\end{equation}
 And the theorem is proved.

\end{Proof}
\end{theorem}
It is obvious that equation~(\ref{modulus1}) is the analogue of equation~(\ref{circle3}), hence on the basis of  theorem (2), the hodograph is a circle of parametric equations

  \begin{equation}\left\{\begin{array}{lr} V_x=-a\dot\theta_b\,\sin\theta&~\\
  V_y=(c+a\cos\theta)\,\dot\theta_b&~\\
  \end{array}\right.~\label{parametric}\end{equation}
   So, in accordance with equation~(\ref{modulus1}), The hodograph of the motion (figure~\ref{Hodo})  is  an off-center circle of radius $V_c=a\,\dot\theta_b$ and center $(V_0=c\,\dot\theta_b,\phi_0=\displaystyle\frac{\pi}{2})$\,, traced in the velocity space by making use of its parametric equations~(\ref{parametric})\,.

The hodograph of a body in uniform rectilinear motion is a fixed point in the velocity space. Correlatively, the existence of the hodograph curve is a manifestation of departure from uniform rectilinear motion. If two or more motions present the same hodograph, then, these motions undergo the same deviation from uniform rectilinear motion. It appears from equation~(\ref{velocityy}) that the motion of planets in elliptic orbits is a combination of a uniform rectilinear part represented by the component $c\,\dot\theta_b\,\bm{\hat{e}_y}$ and a uniform circular part represented by the component $a\,\dot\theta_b\,\bm{\hat{e}_\theta}$\,. It is evident that the deviation from uniform rectilinear motion that the planet undergoes in elliptic motion is restricted to the uniform circular part, and that this deviation is exactly the same as that it would do had its motion been uniform circular at an earlier stage. Therefrom, one can speak about the uniqueness of the hodograph of planets vis \`a vis the changes in the eccentricity of their elliptic trajectories influenced by the decrease in the mass of the star about which they revolve. In a paper~\cite{adel} published in 2019, I proved that a uniform circular motion of a spaceship around a planet consists of an infinite number of successive infinitesimal free falls, a fact that explains the absence of the sensation of gravity aboard a spaceship revolving a planet in a uniform circular motion. The same reasoning applies here, so, one can attribute the absence of the sensation of the gravity of stars on planets, to the sameness of the deviation from uniform rectilinear motion for elliptic and circular trajectories.\\
To recapitulate, the hodograph of a planet orbiting a star is invariant under mass dissipations occurring in the star.

\vspace{-.25cm}

\section*{Conclusion}
\vspace{-.25cm}
\noindent As a matter of fact, all credit goes to  Newton who was the first to allude to  relation~(\ref{Newton}) in Principia Mathematica by saying literally~\cite{Newtonn}$\colon$

\begin{quote}``If a body $P$, by means of a centripetal force tending to any given point $R$, move in the perimeter of any given conic section whose center is $C$; and the law of centripetal force is required$\colon$ draw $CG$ parallel to the radius $RP$\,, and meeting the tangent $PG$ of the orbit in $G$; and the force required (by Cor.1, and Schol. X, and Cor.3, Prop.VII) will be as $\displaystyle\frac{CG^3}{RP^2}$\,.''
\end{quote}
Equation~(\ref{Newton}) giving the expression of the central force acting on a body in elliptic motion around a center of force is in complete agreement with what Newton predicted. Nevertheless, it constitutes a step in advance by realizing that, the point G to which Newton referred and which is called $Z$ in  figure \ref{geometry} belongs to the principal circle, and as such we recognize that his $CG$ is nothing but the length of the semi major axis $OZ=a$, furthermore, it provides an explicit formula to the value of the centripetal force in terms of the geometric parameters of the trajectory and the mass of the planet i.e. an equality and not a proportionality. In other words the missing constant in Newton's prediction turned out to be $m\,(1-\epsilon^2)\,\dot\theta_b^2$\,.\\

\end{document}